\documentclass[floatfix,showpacs,prl,aps,a4paper,twocolumn]{revtex4}

 \usepackage{bm}
 \usepackage{amsmath}
 \usepackage{amssymb}
 \usepackage{latexsym}
 \usepackage{amsfonts}
 \usepackage{epsfig}
 \usepackage{subfigure}
 \usepackage{color}
 \definecolor{darkblue}{rgb}{0,0,.5}
 \usepackage[linktocpage, colorlinks=true, linkcolor=darkblue, citecolor=darkblue]{hyperref}
 \usepackage[all]{hypcap}

\newcommand{\C}[1]{{\cal{#1}}}

\newcommand{\mb}[1]{\mathbb{#1}}
\newcommand{\ov}[1]{{\overline{#1}}}

\begin{document}

\title{Thermodynamics of a physical model implementing a Maxwell demon}

\author{Philipp Strasberg$^1$}
\author{Gernot Schaller$^1$}
\author{Tobias Brandes$^1$}
\author{Massimiliano Esposito$^2$}
\affiliation{
$^1$ Institut f\"ur Theoretische Physik, Technische Universit\"at Berlin, Hardenbergstr. 36, D-10623 Berlin, Germany\\
$^2$ Complex Systems and Statistical Mechanics, University of Luxembourg, L-1511 Luxembourg, Luxembourg}

\begin{abstract}
We present a physical implementation of a Maxwell demon which consists of a conventional single electron transistor (SET)
capacitively coupled to another quantum dot detecting its state. Altogether, the system is described by stochastic thermodynamics. 
We identify the regime where the energetics of the SET is not affected by the detection, but where its coarse-grained 
entropy production is shown to contain a new contribution compared to the isolated SET.
This additional contribution can be identified as the information flow generated by the ``Maxwell demon'' feedback in an idealized limit.
\end{abstract}

\pacs{
05.70.Ln,  
05.40.-a   
05.60.Gg,  
}

\maketitle


For more than a century the thermodynamic implications of various types of ``intelligent interventions''  
(e.g. feedbacks) on the microscopic degrees of freedom of a system have intrigued scientists~\cite{Leff}. 
A Maxwell demon for example can be thought of as a hidden idealized mechanism which is able to modify the second law of 
thermodynamics (entropy balance) but without modifying the first law (energy balance). 
Such a demon would thus be able to heat a hot reservoir while cooling down a cold reservoir without using any 
additional energy, which clearly breaks the traditional formulation of the second law of thermodynamics.

Nowadays, our ability to manipulate small devices has drastically increased, and what used to be unrealistic thought experiments 
have become real experiments~\cite{Leigh, ToyabeSagawaUedaNP10, LutzCiliberto12NP}. 
In parallel to that, significant progress in understanding the nonequilibrium thermodynamics of small systems has been 
achieved~\cite{JarzynskiRev11, EspositoReview, HanggiFTRMP11}.
This is particularly true for systems described by Markovian stochastic dynamics where a consistent theoretical 
framework, called stochastic thermodynamics, has emerged and has proven very useful to study fluctuations and 
efficiencies of systems driven far from equilibrium~\cite{Seifert12Rev, EspoVdB10_Da, EspoVdB10_Db}.
Quite naturally, recent studies have started considering the thermodynamic description of systems subjected to different types
of feedbacks~\cite{VandenBroeck07,SagawaUedaPRL08, SagawaUedaPRL09, SagawaUedaPRL10, Segal10, Horowitz10, HorowitzParrondo11, 
HorowitzParrondo11NJP, EspoVdB_EPL_11, AverinPekola11, SeifertEPL11, SeifertBauerAbreu12, SagawaUeda12, JarzynskyPNAS12, EspositoSchaller12, HorowitzSagawaParrondo12}. 

To understand the thermodynamic behavior of feedback controlled systems it is important to include the \emph{information} 
generated or used by the feedback~\cite{SagawaUeda12, EspositoSchaller12}.
Nevertheless, every feedback scheme has to be implemented physically and the natural question which arises -- and which has not been answered yet -- is 
under which circumstances this physical implementation appears as pure information entering the thermodynamic description.
This is the object of the present letter.

\begin{figure}[ht]
\includegraphics[width=0.3\textwidth,clip=true]{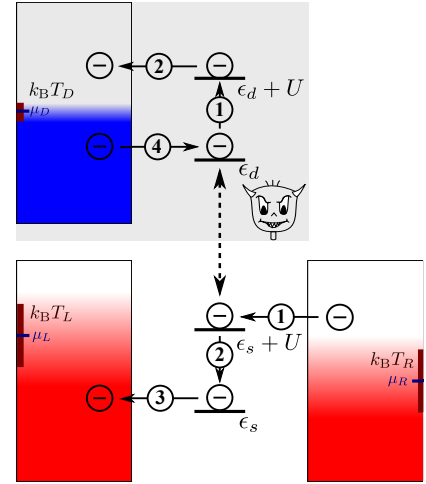}
\caption{\label{fig sketch model}(Color online)
The shaded region constitutes the demon, the contained dot $d$ couples capacitively to the lower dot $s$ via Coulomb repulsion $U$.
The latter is placed in a conventional SET-setup, which can be accessed experimentally, whereas the demon remains hidden.
The trajectory (with initial state $\rho=1$ and $\sigma=E$) marked by the steps $1-4$, 
where an electron is transferred against the bias through the SET, becomes likely in the Maxwell demon limit. 
Note that also the tip of a scanning tunneling microscope could represent the detector dot $d$ with its reservoir, such
that in principle SET and demon need not share the same substrate.
}
\end{figure}

Our model consists of two single level quantum dots interacting capacitively via a Coulomb repulsion $U$ and 
additionally coupled to thermal reservoirs as depicted in Fig.~\ref{fig sketch model}.
No electrons can be transferred between the dots. The corresponding system Hamiltonian is given by  
$H = \epsilon_d c_d^\dagger c_d + \epsilon_s c_s^\dagger c_s + U c_d^\dagger c_d c_s^\dagger c_s$ with
fermionic operators annihilating electrons on the system dot ($c_s$) and the detecting dot ($c_d$).
The dots are weakly coupled to ideal reservoirs $\nu \in \{D,L,R\}$ at temperature $T_\nu = 1/\beta_\nu$ ($k_B\equiv1$) 
and chemical potential $\mu_\nu$. 
As shown in Fig.~\ref{fig sketch model}, dot $s$ is coupled to the two reservoirs 
$L$ and $R$ and constitutes a usual SET, while dot $d$ is coupled to reservoir $D$.
One can show that dot $d$ with reservoir $D$ can be tuned to ``detect'' the state of the SET~\cite{BrandesSchallerPRB10} 
and will eventually constitute the Maxwell demon.
We denote the four eigenvectors of the two coupled dots by $|\rho\sigma\rangle = |\rho\rangle_{d}\otimes|\sigma\rangle_{s}$ 
where $\rho\in\{0,1\}$ and $\sigma\in\{E,F\}$ denote the states of the dot $d$ and $s$, respectively, 
which are either empty (0 or $E$) or filled (1 or $F$).
For weak dot-reservoir interaction, the time evolution of the coupled dots density matrix $\rho$ can be shown to be governed by 
a Markovian master equation $\frac{d}{dt}\rho = \C W\rho$, which simply yields a rate equation for 
the probabilities $p_{\rho\sigma}$ to be in the eigenstate $|\rho\sigma\rangle$.
In the ordered basis $(p_{0E},p_{1E},p_{0F},p_{1F})$ the rate matrix reads
\begin{equation}
\label{eq full dynamics}
 \C W = \left(\begin{array}{cc}
               \C W_{EE}	&	\C W_{EF}	\\
               \C W_{FE}	&	\C W_{FF}	\\
              \end{array}\right)
\equiv \sum_\nu \C W^{(\nu)}
\,.
\end{equation}
The superscript $\nu$ denotes transitions triggered by the respective reservoir $\nu\in\{D,L,R\}$.
The two blocks
\begin{align}
 \C W_{EE} =& \left(\begin{array}{cc}
                    -\gamma_D-\gamma_L-\gamma_R	&	\ov\gamma_D				\\
                    \gamma_D			&	-\ov\gamma_D-\gamma_L^U-\gamma_R^U	\\
                   \end{array}\right),	\\
 \C W_{FF} =& \left(\begin{array}{cc}
                    -\gamma_D^U-\ov\gamma_L-\ov\gamma_R	&	\ov\gamma_D^U					\\
                    \gamma_D^U				&	-\ov\gamma_D^U-\ov\gamma_L^U-\ov\gamma_R^U	\\
                   \end{array}\right)
\end{align}
denote the dynamics of the dot $d$ when dot $s$ is empty or filled, respectively, while the blocks
\begin{align}
\C W_{EF} =& {\rm Diag}(\ov\gamma_L+\ov\gamma_R, \ov\gamma_L^U+\ov\gamma_R^U)\,,\\
\C W_{FE} =& {\rm Diag}(\gamma_L+\gamma_R,\gamma_L^U+\gamma_R^U)
\end{align}
denote the transitions between an empty and filled dot $s$ which can occur when dot $d$ is either empty or filled.
The rates are given by $\gamma_\nu \equiv \Gamma_\nu f_\nu$, $\gamma_\nu^U \equiv \Gamma_\nu^U f_\nu^U$, $\ov\gamma_\nu \equiv \Gamma_\nu(1-f_\nu)$ and 
$\ov\gamma_\nu^U\equiv \Gamma_\nu^U (1-f_\nu^U)$ with electronic tunneling rates
$\Gamma_\nu>0$ and $\Gamma_\nu^U>0$ and Fermi functions $0 < f_\nu,f_\nu^U < 1$ with $\nu\in\{D,L,R\}$. 
Note that $\Gamma_\nu \neq \Gamma_\nu^U$ requires to go beyond the common wide-band approximation.
The Fermi functions are evaluated at the respective transition energies of the SET dot $f_{L/R} \equiv 1/[\exp(\beta_{L/R}(\epsilon_s-\mu_{L/R}))+1]$, 
$f_{L/R}^U \equiv 1/[\exp(\beta_{L/R}(\epsilon_s+U-\mu_{L/R}))+1]$ and the detector dot $f_D \equiv 1/[\exp(\beta_D(\epsilon_d-\mu_D))+1]$, 
$f_D^U \equiv 1/[\exp(\beta_D(\epsilon_d+U-\mu_D))+1]$, respectively. From now on we will assume $\beta_L=\beta_R=\beta$.
The present rate equation description is valid when $\beta_\nu \Gamma_\nu^{(U)} \ll 1$ but may remain qualitatively correct even outside this range.

This system is consistently described by stochastic thermodynamics. 
At steady state the entropy production of the full system reads 
\begin{equation}\label{TotalEPbis}
 \dot S_i = \sum_\nu \sum_{\rho,\rho',\sigma,\sigma'} W_{\rho\sigma\rho'\sigma'}^{(\nu)} p_{\rho'\sigma'} \ln\frac{W_{\rho\sigma\rho'\sigma'}^{(\nu)}}{W_{\rho'\sigma'\rho\sigma}^{(\nu)}} \ge 0 \;.
\end{equation}
It can be rewritten after some algebra as the sum of force-flux terms associated to matter and energy transfers
\begin{equation}\label{TotalEP}
 \dot S_i = \beta (\mu_L-\mu_R) I_S + (\beta_D-\beta) I_E \ge 0,
\end{equation}
where $I_S = \gamma_L p_{0E} - \ov\gamma_L p_{0F} + \gamma_L^Up_{1E} - \ov\gamma_L^U p_{1F}$ is the stationary electronic particle current flowing 
from reservoir $L$ to $R$ through the SET and $I_E = U (\gamma_D p_{0E} - \ov\gamma_D p_{1E})$ is the energy current entering 
the reservoir $D$ due to the interaction between dot $d$ and the SET. 
The matter current associated to the detector bath $D$ vanishes since there is no particle exchange between dot $d$ and dot $s$. 

Using usual techniques~\cite{AndrieuxGaspard07a,EspositoReview}, it is possible to show that 
the following fluctuation theorem for the entropy production is satisfied: 
\begin{equation}
 \lim_{t\rightarrow\infty} \frac{p_{+n_S,+n_D}(t)}{p_{-n_S,-n_D}(t)} = e^{\beta(\mu_L-\mu_R)n_S+(\beta_D-\beta)U n_D}\,,
\end{equation}
where $p_{n_S,n_D}(t)$ denotes the probability of having $n_S$ electrons traversing the system from left to right together
with net $n_D$ electrons entering dot $d$ at energy $\epsilon_d$ and leaving it at energy $\epsilon_d+U$, altogether leading
to a net energy transfer of $U \cdot n_D$ into reservoir $D$ after time $t$.
So far, our system is thus only a conventional thermoelectric device in which the thermal gradient may be used to generate an 
electronic current through the SET against the bias.
Related models have been considered in~\cite{EspoRuttCleuPRB,BrandesSchallerPRB10, ButtikerSanchez10PRL, ButtikerSanchez11PRL, EspositoCuetaraGaspard11}.

\begin{figure}[ht]
\includegraphics[width=0.35\textwidth,clip=true]{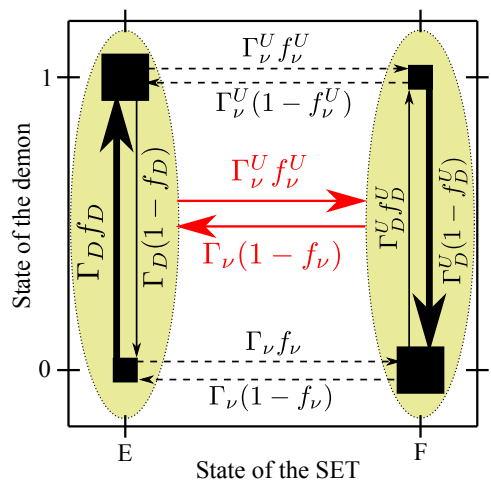}
\caption{\label{fig mesostates}(Color online)
Visualization of the dynamics in the full (black arrows) and coarse-grained (red arrows) state space. 
The four states are denoted by black squares with their size being proportional to the occupation probability
near the sought-after limit.
The two mesostates of the SET ($E$ or $F$) are composed by the shaded regions. 
They each contain two possible states of the detector dot $d$ which are connected by fast transitions. 
The red labels denote transition rates between the two SET mesostates in the error-free limit ($f_D \to 1$, $f_D^U\to 0$ and 
$\Gamma_D/\Gamma_\alpha^{(U)}\to\infty$ for $\alpha\in\{L,R\}$).
}
\end{figure}

We now assume that our experimental setup allows us to detect electron transfers in the SET (e.g. counting statistics experiments) 
but does not provide any information about the existence of the demon (i.e., dot $d$ and reservoir $D$). 
The observed SET states $\sigma\in\{E,F\}$ thus constitute two coarse-grained ``mesostates'' with probabilities 
$p_\sigma=p_{0\sigma} + p_{1\sigma}$, see Fig.~\ref{fig mesostates}. 
Moreover, we denote by $\mb{P}_{\rho|\sigma}=p_{\rho\sigma}/p_\sigma$ the conditional stationary 
probability to find dot $d$ in state $\rho$ if the state of the SET is $\sigma$.
Each of these probabilities can be explicitly 
calculated from the analytical steady-state solution of Eq.~(\ref{eq full dynamics}).
It is straightforward to see that the exact coarse-grained dynamics of the SET may formally be written as
\begin{equation}\label{Erateeff}
\dot{p}_\sigma = \sum_{\sigma'} V_{\sigma\sigma'} p_{\sigma'}
\end{equation}
with ``rates'' $V_{\sigma\sigma'} = \sum_\nu V_{\sigma\sigma'}^{(\nu)} = \sum_\nu \sum_{\rho\rho'} W_{\rho\sigma,\rho'\sigma'}^{(\nu)} \mb{P}_{\rho'|\sigma'}$.

{\it Fast demon}: Since we want dot $d$ and reservoir $D$ to ultimately constitute a Maxwell demon, 
we are now going to assume that the dynamics of the demon is much faster than the SET dynamics 
$\Gamma_D=\Gamma_D^U \gg \max\{\Gamma_L^{(U)},\Gamma_R^{(U)}\}$. 
As expected, in the extreme case $\Gamma_D/\max\{\Gamma_{L/R}^{(U)}\}\rightarrow\infty$, 
the conditional probabilities equilibrate instantaneously with respect to the reservoir $D$ 
\begin{equation}
 \begin{split}\label{eq conditional probabilities under AE}
  \mb{P}_{0|E}	&\to \frac{\ov\gamma_D}{\gamma_D+\ov\gamma_D} = 1-f_D
\,, ~~~\mb{P}_{1|E} \to 1-\mathbb{P}_{0|E}\,,\\
  \mb{P}_{0|F}	&\to \frac{\ov\gamma_D^U}{\gamma_D^U+\ov\gamma_D^U} = 1-f_D^U
\,, ~~~\mb{P}_{1|F} \to 1-\mb{P}_{0|F}\,,
 \end{split}
\end{equation}
such that Eq.~(\ref{Erateeff}) becomes an ordinary rate equation.
For finite demon temperatures, there will thus always be some finite detection ``error'', which we quantify by $\epsilon_E\equiv\mb{P}_{0|E}/\mb{P}_{1|E}$ 
and $\epsilon_F \equiv\mb{P}_{1|F}/\mb{P}_{0|F}$. 
For instance, when $\epsilon_d=\mu_D-U/2$, these errors will be given by $\epsilon_E=\epsilon_F=\exp(-\beta_D U/2)$.

The entropy production corresponding to the coarse-grained SET dynamics is 
\begin{equation}\label{EffEP}
 \dot S_i^* = \sum_{\nu,\sigma,\sigma'} V_{\sigma \sigma'}^{(\nu)} p_{\sigma'} \ln \frac{V_{\sigma \sigma'}^{(\nu)}}{V_{\sigma' \sigma}^{(\nu)}}
 = A^*_S I_S \ge 0
\end{equation}
with an effective affinity
\begin{equation}\label{Eeffaff}
A_S^* = \ln\left[\frac{\left(\ov\gamma_D\gamma_L+\gamma_D\gamma_L^U\right)\left(\ov\gamma_D^U\ov\gamma_R+\gamma_D^U\ov\gamma_R^U\right)}
{\left(\ov\gamma_D^U\ov\gamma_L+\gamma_D^U\ov\gamma_L^U\right)\left(\ov\gamma_D\gamma_R+\gamma_D\gamma_R^U\right)}\right]\,.
\end{equation}
This coarse-grained entropy production always under-evaluates the full entropy production: 
$\Delta \dot S_i \equiv \dot S_i - \dot S_i^*\geq 0$~\cite{Esposito12}.
Furthermore, it implies an effective fluctuation theorem for the particle counting statistics of the SET 
(e.g. see~\cite{EspositoCuetaraGaspard11})
\begin{equation}\label{EffFT}
 \lim_{t\rightarrow\infty} \frac{p_{+n_S}(t)}{p_{-n_S}(t)} = e^{A^*_S n_S}\,,
\end{equation}
where $p_{n_S}(t)$ is the probability of having $n_S$ electrons transferred from left to right after time $t$.
This demonstrates that the coarse-grained entropy production~(\ref{EffEP}) is a meaningful and measurable quantity characterizing the SET in the fast-demon limit.
Since the entropy production of an isolated SET is $\dot{S}_i^{\rm SET} = \beta(\mu_L-\mu_R) I_S$, the coarse-grained entropy production can be written as 
\begin{equation}\label{EffEPbis}
\dot{S}_i^* = \dot{S}_i^{\rm SET} + I_F \,.
\end{equation}
In ignorance of the physical nature of the demon which has been traced out, $I_F$ must be interpreted as an information current between demon and system which modifies the second law of the SET. 

{\it Fast and Precise demon}: 
Since the demon also needs to be able to reliably discriminate between the two states of the SET, we put $\epsilon_d=\mu_D-U/2$ and further assume that $\beta_D U \gg 1$, which implies $\ov\gamma_D,\gamma_D^U\to 0$ and $\gamma_D,\ov\gamma_D^U\to\Gamma_D$.
As a result the two dots become perfectly correlated, meaning that when the SET gets filled up (resp. emptied) the detecting dot gets emptied (resp. filled up) immediately after. The steady state then reads $p_{1E}=(\ov\gamma_L+\ov\gamma_R)/(\ov\gamma_L+\ov\gamma_R+\gamma_L^U+\gamma_R^U)$, $p_{0F} = 1-p_{1E}$, $p_{0E} = p_{1F} = 0$ and
the effective affinity~(\ref{Eeffaff}) becomes
\begin{equation}
A_S^* = \ln\left[
\frac{f_L(1-f_R)}{(1-f_L) f_R}
\frac{f_L^U f_R}{f_L f_R^U}
\frac{\Gamma_L^U \Gamma_R}{\Gamma_L \Gamma_R^U}\right]\,.
\end{equation}
As a result the information current reads
\begin{equation}\label{Einfcurrent1}
I_F = \ln\left[\frac{f_L^U f_R}{f_L f_R^U}
\frac{\Gamma_L^U \Gamma_R}{\Gamma_L \Gamma_R^U}\right] I_S\,.
\end{equation}
We note however, that this demon is not yet a true Maxwell demon because it is effectively extracting energy from the SET at a rate 
\begin{equation}
I_E = U \frac{(\ov\gamma_L+\ov\gamma_R)(\gamma_L^U+\gamma_R^U)}{\ov\gamma_L+\ov\gamma_R+\gamma_L^U+\gamma_R^U},
\label{ActivityCurr}
\end{equation}
and is thus creating an imbalance of order $U$ between the energy currents at the left and right interface of the SET: $I_E=-I_E^{(L)}-I_E^{(R)}$.
We note that the right hand side of (\ref{ActivityCurr}) equals $U/2$ times the \emph{activity current} in the SET which measures the total number of electron jumps in and out of the SET. This is due to the fact that each change in the mesostates ($E \leftrightarrow F$) instantaneously induces a jump in dot $d$ ($1\leftrightarrow0$) and in the error-free limit no other contribution arises. 
Finally, since $I_E$ remains finite as $\beta_D\to\infty$, the total entropy production diverges in that limit $\dot{S}_i \to \infty$. This means that the demon generates an infinite dissipation $\Delta \dot S_i \to \infty$ to be able to operate in the fast and precise limit.

{\it Maxwell demon}: In order to obtain a true Maxwell demon, we further need to assume that the temperatures of the left and right reservoirs are sufficiently large compared to the capacitive interaction $U$ such that $\beta U \rightarrow 0$. 
In this limit, the energetics of the SET is not affected anymore by the demon since the Fermi functions evaluated at the different energies become equal: 
\begin{equation}
 \lim_{\beta U\rightarrow0} f_\nu^U = f_\nu \ \ \;, \ \ \nu\in\{L,R\} \;.
\end{equation}
The energy imbalance $I_E$ then can be made arbitrarily small in comparison to the SET energy currents since their ratio is of order $U/\epsilon_s$.

The bare rates $\Gamma_\nu^{(U)}$ however could and -- to obtain any nontrivial effect -- should depend on the energy levels of the dots. 
As a result, the information current~(\ref{Einfcurrent1}) becomes
\begin{equation}\label{I_FMaxdemon}
I_F = (\delta_L-\delta_R)I_S,
\end{equation}
where we introduced the feedback parameters $\delta_\nu = \ln \left[\Gamma^U_\nu/\Gamma_\nu\right]$ with $\nu\in\{L,R\}$.
Remarkably, the coarse-grained rate matrix~(\ref{Erateeff}) describing the 
effective SET dynamics now satisfies the modified local detailed balance condition
\begin{equation}
\ln \frac{V_{FE}^{(\nu)}}{V_{EF}^{(\nu)}}= -\beta_{\nu} (\epsilon_{s}-\mu_{\nu}) + \delta_\nu \ \ \;, \ \  \nu\in\{L,R\}.
\label{LDB}
\end{equation}
This result is in perfect agreement with the modified local detailed balance condition introduced 
in~\cite{EspositoSchaller12} to describe Maxwell demon feedbacks within the framework of stochastic thermodynamics.
Alternatively, such a modification of local detailed balance may be generated using a fast feedback control loop~\cite{BrandesSchallerPRB11}.
Now, when $\delta_L\ll 0$ and/or $\delta_R \gg 0$, a trajectory as shown in Fig.~\ref{fig sketch model} becomes highly probable.
\begin{figure}[ht]
\includegraphics[width=0.48\textwidth,clip=true]{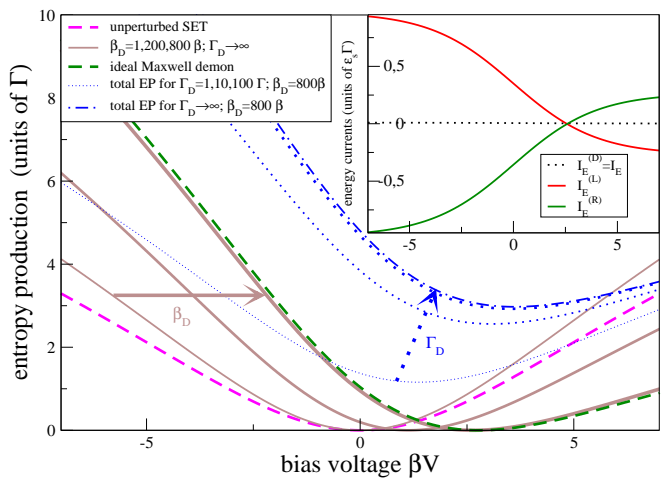}
\caption{\label{fig entropy prod}(Color online) 
The coarse-grained entropy production $\dot S_i^*$ as a function of the SET bias $\beta V$ in the fast demon limit (solid brown curves)
differs from the unperturbed SET case (dashed magenta) in nonequilibrium and moves toward the ideal Maxwell demon limit (dashed green) as the 
temperature of the detector bath (and correspondingly the error of the demon) is lowered ($\beta_D=1\beta,200\beta,800\beta$ with increasing thickness, solid arrow).
The total entropy production always exceeds the coarse-grained one for similar parameters (e.g. dash-dotted blue vs. bold-solid brown for $\beta_D=800$).
Dotted blue curves demonstrate the rapid convergence of the full entropy production to the fast demon limit 
when $\Gamma_D \gg \Gamma$ (dotted arrow).
The SET (solid) and demon (dotted) energy current in the inset (for $\Gamma_D=100\Gamma$ and $\beta_D=800\beta$) demonstrate that the 
relative modification of the SET's first law
is of order $U/\epsilon_s$, which can be made arbitrarily small in comparison to the SET energy currents nearly everywhere.
Other parameters were chosen as
$\Gamma_L=\Gamma_R^U=e^{+\delta} \Gamma$, $\Gamma_L^U=\Gamma_R=e^{-\delta}\Gamma$ with 
$\delta = \ln 2$, $\beta_L=\beta_R=\beta$, $\mu_L = \epsilon_s+V/2$, $\mu_R=\epsilon_s-V/2$, $\beta\epsilon_s = 1$, $\beta U = 0.01$.
}
\end{figure}
Naturally, even without the strict mathematical limits that we discussed, 
we demonstrate in Fig.~\ref{fig entropy prod} that the coarse-grained
entropy production $\dot S_i^*$ approaches the ideal Maxwell demon from~\cite{EspositoSchaller12}. 
Furthermore, the inset demonstrates that for $\epsilon_s \gg U$, the modification of the first law for an isolated SET 
$I_E^{(L)}+I_E^{(R)}=-I_E\approx 0$ is negligible.

We note that even the strict Maxwell demon limit is well-described by our model, 
as all the required inequalities can be simultaneously fulfilled:
$\beta_\nu \Gamma_\nu^{(U)} \ll 1$ (weak coupling), 
$\Gamma_D^{(U)}/\Gamma_{L/R}^{(U)} \gg 1$ (fast measurement),
$\beta_D U \gg 1$ (precise measurement),
$\beta_{L/R} U \ll 1$ (neglect of back-action), and
$U/\epsilon_s \ll 1$ (preservation of SET energy currents).
For example, the curves with finite $\Gamma_D$ in Fig.~\ref{fig entropy prod} 
only require a sufficiently small base tunneling rate $\Gamma$.

We now turn to the interpretation of our results. 
We have seen that the true entropy production of the system and the detector, $\dot{S}_i$~(\ref{TotalEP}), diverges 
when the detector performs a perfect feedback (i.e., infinitely fast and precise) on the system. 
This is conceptually very important, but of low practical interest to assess how effective is the feedback 
in generating gains at the system level. To do so one has to simply discard the demon dissipation and focus 
on the coarse-grained entropy production $\dot{S}_i^*$~(\ref{EffEPbis}) which characterizes the entropy 
production of the system subjected to the information current $I_F$ generated by the feedback. 
Using this entropy production one can study the thermodynamic efficiency with which a negative 
information current can be used to generate various processes on the system (e.g. to transport 
electrons against the bias or to cool a cold reservoir).
In contrast, the efficiencies of the thermoelectric device as a whole may be drastically lower, 
since in that case the total entropy production $\dot{S}_i$ should be considered. 

{\it Conclusion}
To the best of our knowledge, this letter establishes for the first time the precise connection between 
the complete thermodynamic description of a Maxwell demon model and the system it is acting on.
In particular, we have identified the effective level of description of the system where the demon manifests 
itself solely through an information flow modifying the second law. 
Furthermore, by showing that the effective entropy production is only a piece of the total 
entropy production of the joined system, we provide a rigorous support for the generic claim that a system 
subjected to a ``Maxwell demon'' is an idealization which neglects the dissipation associated to the 
implementation of the demon mechanism.

{\it Acknowledgments} 
Financial support by the DFG (SCHA 1646/2-1, SFB 910, and GRK 1558) and the National Research Fund, 
Luxembourg (project FNR/A11/02) is gratefully acknowledged.


\end{document}